\documentclass[prl,amsmath,amssymb,showpacs,reprint,superscriptaddress]{revtex4-1}
\usepackage{graphicx}
\begin{document}
\title{High accuracy measure of atomic polarizability in an optical lattice clock}

\author{J.\ A.\ Sherman}\email{jeff.sherman@nist.gov}
\affiliation{National Institute of Standards and Technology, 325 Broadway, Boulder, Colorado 80305, USA}
\author{N.\ D.\ Lemke}
\author{N.\ Hinkley}
\affiliation{National Institute of Standards and Technology, 325 Broadway, Boulder, Colorado 80305, USA}
\affiliation{University of Colorado, Department of Physics, Boulder, Colorado 80309, USA}
\author{M.\ Pizzocaro}
\affiliation{Politecnico di Torino, Corso duca degli Abruzzi 24, 10125 Torino, Italy}
\author{R.\ W.\ Fox}
\author{A.\ D.\ Ludlow}
\author{C.\ W.\ Oates} 
\affiliation{National Institute of Standards and Technology, 325 Broadway, Boulder, Colorado 80305, USA}

\date{\today}
\begin{abstract}
Despite being a canonical example of quantum mechanical perturbation theory, as well as one of the earliest observed spectroscopic shifts, the Stark effect contributes the largest source of uncertainty in a modern optical atomic clock through blackbody radiation.  By employing an ultracold, trapped atomic ensemble and high stability optical clock, we characterize the quadratic Stark effect with unprecedented precision.  We report the ytterbium optical clock's sensitivity to electric fields (such as blackbody radiation) as the differential static polarizability of the ground and excited clock levels $\alpha_\text{clock} = 36.2612(7)~\text{kHz (kV/cm)}^{-2}$.  The clock's fractional uncertainty due to room temperature blackbody radiation is reduced an order of magnitude to $3 \times 10^{-17}$.
\end{abstract}

\keywords{polarizability, Stark, optical clock, ytterbium, blackbody}
\pacs{32.10.Dk, 32.60.+i, 06.20.fb,44.40.+a}
\maketitle

\begin{figure}[b]  
\includegraphics[width=3.3in]{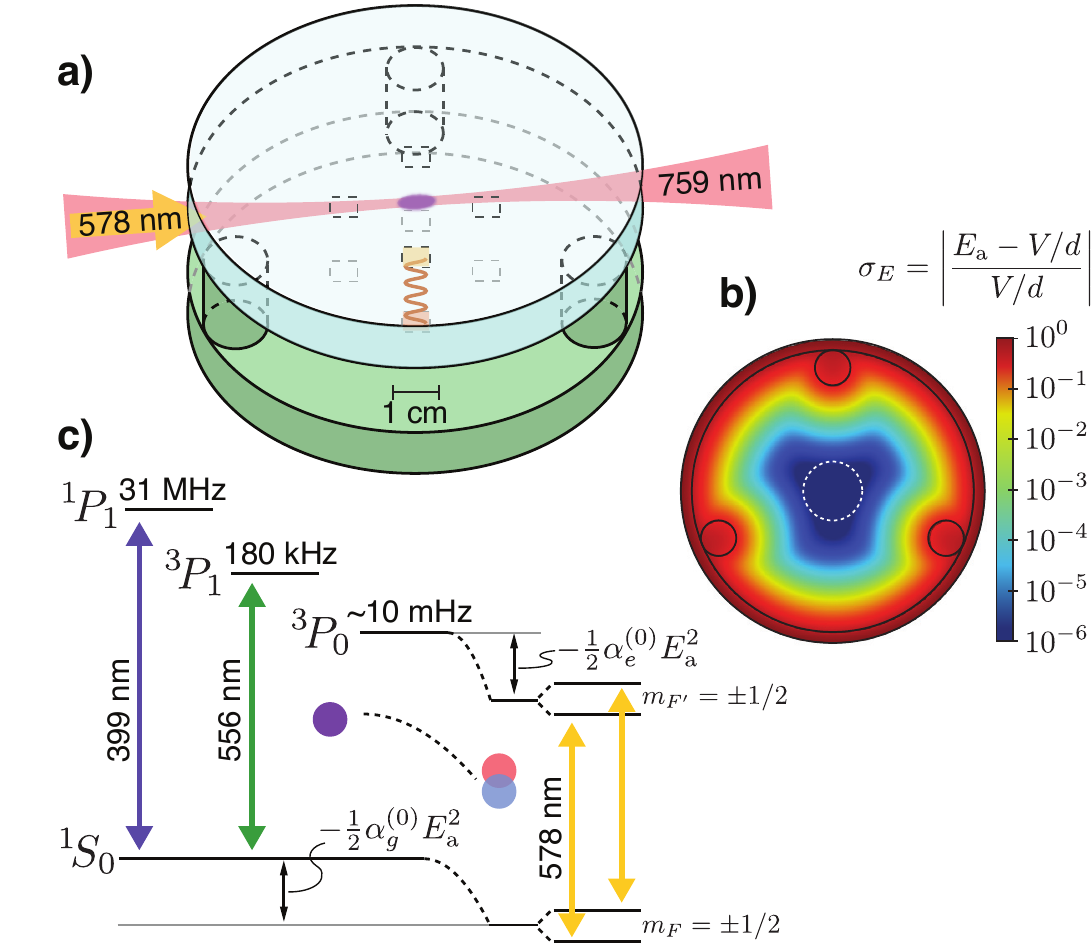}
\caption{a) A scale drawing of transparent conductive electrodes surrounding a one-dimensional optical lattice of ytterbium atoms.  Four pairs of metallic pads (one of which is highlighted) allow \emph{in situ} interferometric measurement of the electrode spacing $d$. b) Deviations $\sigma_E$ of the applied electric field $E_\text{a}$ from the ideal field created by infinite-planar electrodes become negligible in the central ~1 cm region (dashed white circle). c) Relevant energy levels, laser transition wavelengths, and linewidths ($\Gamma/2 \pi$) of $^{171}$Yb.  Clock states $^1 \!S_0$ and $^3 \! P_0$ are shown Stark-shifted by $\vec{E}_a$.}
\label{fig:levels}
\end{figure}
An atom immersed in an electric field $\vec{E}_\text{a}$ becomes \emph{polarized}---the electronic wave-function is stretched into alignment with the field.  Generally, energies of the lowest-lying electronic quantum states $| i \rangle$ are reduced by $- \tfrac{1}{2} \alpha_i^{(0)} E_\text{a}^2$ (see Fig.~\ref{fig:levels}) where $\alpha_i^{(0)}$ is termed a state's static polarizability~\cite{mitroy2010theory}.  The scaling of the Stark effect is quadratic because $\vec{E}_\text{a}$ is responsible for inducing, and also interacting with, an atomic dipole moment.

Neutral atom lattice clocks~\cite{katori2003ultrastable} employ $10^3$--$10^5$ ultracold alkaline-earth atoms tightly confined in an optical standing wave potential so their intrinsically narrow, largely imperturbable, $^{1}\! S_0 \leftrightarrow {}^{3}\! P_0$ transitions~\cite{porsev2004possibility} may establish stable and accurate frequency and time references~\cite{udem2002optical,hollberg2005optical}.  In analogy to a pendulum's oscillation slowing due to thermal expansion, a ytterbium lattice clock slows when electrically stretched, or polarized, by thermal blackbody radiation (BBR) fields~\cite{lemke2009spin,porsev2006multipolar}.  This phenomenon has been measured in the cesium fountain primary standard~\cite{simon1998measurement, angstmann2006frequencycesium} and other optical transitions~\cite{mitroy2010theory,rosenband2006blackbody,li1996stark}.  

No shield at finite temperature protects a clock atom from the time varying electric field of BBR, the electromagnetic energy absorbed and re-emitted by all matter in thermal equilibrium according to the Stefan-Boltzmann law~\cite{landau5lifshitz}.  Inside a hollow shell of opaque matter (a \emph{blackbody}), the time-averaged electric field intensity depends strongly on temperature:
\begin{align}
\langle E^2 \rangle_T 	&= \frac{1}{2} \int_0^\infty \underbrace{\frac{8 \alpha^3}{\pi} \frac{\omega^3 \, d \omega}{e^{\omega/ k_\text{B} T}-1}}_{E^2(\omega)\, d\omega} = \frac{4 (\alpha \pi)^3}{15}(k_\text{B} T)^4, \\ 
					&\approx \left( 8.319 \text{ V/cm} \right)^2 \left( \frac{T}{300 \text{ K}} \right)^4,  \nonumber
\end{align}
where $\alpha \approx 1/137$ is the fine-structure constant, $k_\text{B}$ is Boltzmann's constant, and $T$ is the blackbody's absolute temperature~\cite{angstmann2006frequency}.  Near room temperature, the spectrum of radiation is peaked strongly near $9.6~\mu$m---invisible to the eye and also far detuned from strong electronic transitions in ytterbium.

The ytterbium clock frequency ($\nu \approx 518$~THz) is shifted by the net BBR Stark effect of the two clock states, which can be expressed as
\begin{equation}
\label{eq:pol}
\Delta \nu_\text{BBR} = -\frac{1}{2} \left( \alpha_e^{(0)} - \alpha_g^{(0)} \right) \langle E^2 \rangle_T \left[1 + \eta_\text{clock}(T) \right],
\end{equation}
where $\alpha_{g,e}^{(0)}$ are the static polarizabilities of ground and excited states ($^1\! S_0$ and $^3 \! P_0$, respectively), and $\eta_\text{clock}(T)$~\cite{porsev2006multipolar} is a small computed parameter accounting for the dynamic aspect of the BBR field~\footnote{State specific parameters ($\eta_g \approx 0, \eta_p = 0.007$) computed in \cite{porsev2006multipolar} help to define $\eta_\text{clock} \equiv (\eta_e \alpha_e^{(0)} - \eta_g \alpha_g^{(0)}) / \alpha_\text{clock}$.}.  At room temperature, $\eta_\text{clock}(300 \text{ K}) = 0.0145(15)$ is a small contribution to $\Delta \nu_\text{BBR} \approx -1.3$~Hz.  More significantly, knowledge of $\alpha_\text{clock} \equiv \alpha_e^{(0)} - \alpha_g^{(0)}$ is theoretical, and limited to 10\% accuracy due to the complexity of this many-electron atom~\cite{dzuba2010dynamic,porsev1999electric,porsev2006multipolar}.

To measure $\alpha_\text{clock}$, and reduce the clock's uncertainty due to BBR, we fitted electrodes~\cite{hunter1992precise} to an existing ytterbium clock apparatus~\cite{lemke2009spin}.  A voltage $V$ on ideal electrodes spaced by $d$ in vacuum creates an electric field $E_\text{a} = V/d$, shifting the clock transition by $\Delta \nu = -\tfrac{1}{2} \alpha_\text{clock} (V/d)^2$.   Deviations from this infinite-parallel plane capacitor model are bounded at the $10^{-6}$ (1~ppm) level by designing a large electrode diameter-to-spacing ratio, ensuring a high degree of parallelism, and centering the atoms radially within the electrodes.  Perturbations due to dielectric and conducting mounting structure contribute similar amounts of field uncertainty.

The electrodes, shown in Fig.~\ref{fig:levels}, are comprised of rigidly spaced parallel fused silica plates ($101.6(1)$~mm in diameter) featuring a transparent conductive coating on the inner surfaces. The outer surfaces are anti-reflection coated for all relevant laser wavelengths. The electrode separation $d = 15.03686(8)$~mm is maintained by three fused silica rods bonded 45(1)~mm from the center axis with hydroxide catalysis~\cite{gwo2001ultra,reid2007influence}.  $d$ is determined interferometrically by measuring (\emph{in situ}) the free-spectral-range $\nu_\text{fsr} = c/2d$ of planar etalons formed by 90\%-reflective metallic pads (33~nm gold on 2~nm chromium) deposited over the 0.3~nm indium-tin-oxide inner electrode faces~\cite{hunter1992precise}.  Each 6~mm square pad is offset 28~mm from center.  An external cavity diode laser is tuned by 17 THz around $\lambda_\text{p} = 766$~nm to observe $\sim 10$ of $N_\text{f} \approx 1700$ etalon transmission peaks.  Each transmission feature, located with a wavelength meter to $\pm 50$~MHz, has a linewidth of 500~MHz, consistent with a finesse $\mathcal{F} \approx 20$.  Spacing the observed peaks logarithmically allows an efficient least-squares determination of $\nu_\text{fsr}$.  Systematic wavelength and fringe-center inaccuracy is largely divided down by $N_\text{f}$.  Uncertainty in the metal pad thickness contributes $3 \times 10^{-7}$ uncertainty to $d$.  Gold's index of refraction ($n_r \approx 0.16$)~\cite{schulz1954optical} varies gently about $\lambda_\text{p}$; variations in $\nu_\text{fsr}$ with fringe index due to mirror phase shifts contribute an error of $1.2 \times 10^{-6}$.  Stray etalons add a similarly sized line-pulling error. Electrode parallelism is constrained by the measured finesse as well as spatially independent pad-pair measurements of $d$.

Thin strips of silver-loaded epoxy join insulated wires to electrode perimeters.  Two wires are redundantly bonded to each electrode to establish the sheet resistance of the conductive layer ($R_\text{ito} = 3 \text{ k}\Omega$).  The parasitic resistance between the electrodes and the grounded vacuum structure is $R_\text{leak} = 316(9)$~G$\Omega$ ($R_\text{leak} = 14.3(5)$~G$\Omega$) below (above) an observed field-emission threshold occurring near 800~V.  Leads to the electrodes have current-limiting resistances of $10$~k$\Omega$ each;  the worst-case $E_\text{a}$ error from voltage division is $1.4 \times 10^{-6}$.  We constructed a regulated voltage source producing 100~V--1050~V with $1.0 \times 10^{-6}$ instability over 1~s--1000~s~\cite{horowitz1989art}.  

A clock interrogation cycle (360~ms) begins with slowing, cooling, and trapping $^{171}$Yb from an atomic beam with a magneto-optical trap operating first on the $^{1}\! S_0 \leftrightarrow {}^1 \! P_1$ transition (399~nm), then on the narrower $^{1}\! S_0 \leftrightarrow {}^3 \! P_1$ transition (556~nm).  Ultracold atoms (10~$\mu$K) are confined by an optical lattice (1D) at the so-called \emph{magic} wavelength near $759$~nm resulting in no net ac-Stark shift between $^{1}\! S_0$ and $^{3}\! P_0$~\cite{porsev2004possibility,barber2008optical}.  Atoms are optically pumped into one nuclear-spin state ($m_F = \pm 1/2$) of $^{1}\! S_0$. A $\pi$-polarized pulse (100~ms) of resonant 578~nm laser radiation coherently excites one of two $\Delta m_F = 0$ transitions (split by 500~Hz by an applied magnetic field $\vec{B}$) to the long-lived $^3 \! P_0$ state (see Fig.~\ref{fig:levels}).  A series of laser pulses converts the resulting clock state populations into fluorescence signals which are then normalized against atom number fluctuations~\cite{lemke2009spin}.

\begin{figure}
\includegraphics[width=3.5 in]{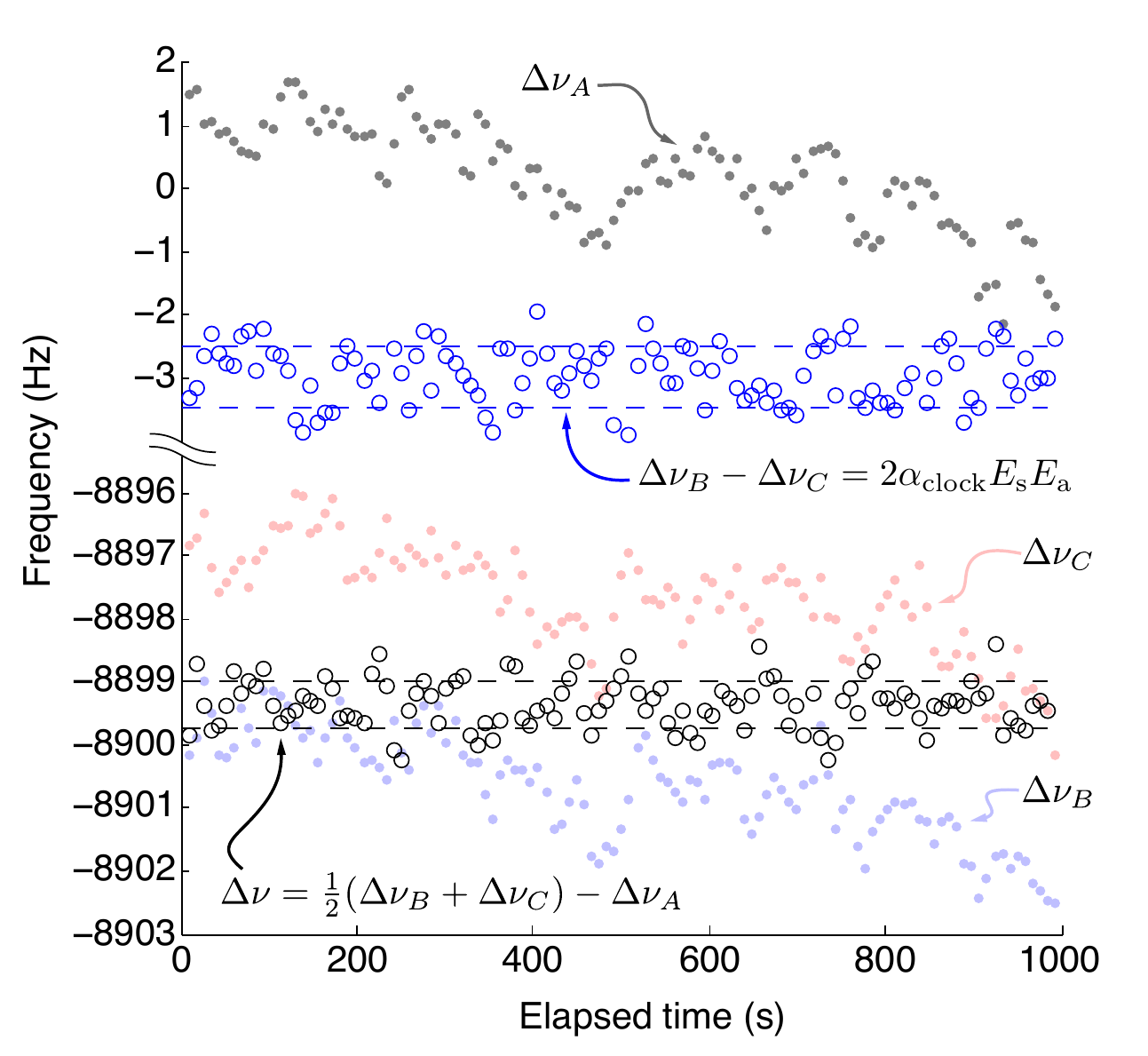}
\caption{Integrated clock laser error signals are shown for $E_\text{a} \approx 700.6$~V/cm under three interleaved conditions:  both electrodes grounded (`A', grey dots), and each electrode at high voltage with the other grounded (`B' and `C', blue and red dots).  Laser drift, common to these signals, is removed from the quadratic Stark shift signal (black open circles, see Eq.~\ref{eq:starkShift}) and the stray field signal $\Delta \nu_B - \Delta \nu_C$ (blue open circles).  Dashed lines show standard deviations for these signals.  A passive feed-forward linear drift canceler reduced clock laser drift to 4~Hz over this 1000~s data run.}
\label{fig:dataPlot}
\end{figure}

The stabilized 578~nm laser is independently locked to the atomic transition under three interleaved conditions: both electrodes grounded (condition `A'), and each electrode at high voltage with the other grounded (`B' and `C').  Each condition, mediated by opto-coupled reed-relays, lasts $\tau_\text{v} = 2.9$~s, during which the clock laser maintains a fractional frequency stability approaching $3 \times 10^{-16}$~\cite{jiang2011making}.  In each period $\tau_\text{v}$, two interrogations are performed on each side of both nuclear-spin spectroscopic features.  Slow laser drifts are common to all three integrated error signals, $\Delta \nu_A$, $\Delta \nu_B$, and $\Delta \nu_C$ (see Fig.~\ref{fig:dataPlot}).  The quadratic Stark shift is 
\begin{equation}
\Delta \nu = \tfrac{1}{2}(\Delta \nu_B + \Delta \nu_C ) - \Delta \nu_A. 
\label{eq:starkShift} 
\end{equation}
The total Allan deviation~\cite{howe2000total} is used to determine the statistical uncertainty of $\Delta \nu$.  For presented data, $\vec{E}_\text{a} \parallel \vec{E}_\text{lattice}$ (both were perpendicular to $\vec{B}$), though other configurations were examined.

\begin{figure} 
\includegraphics[width=3.5in]{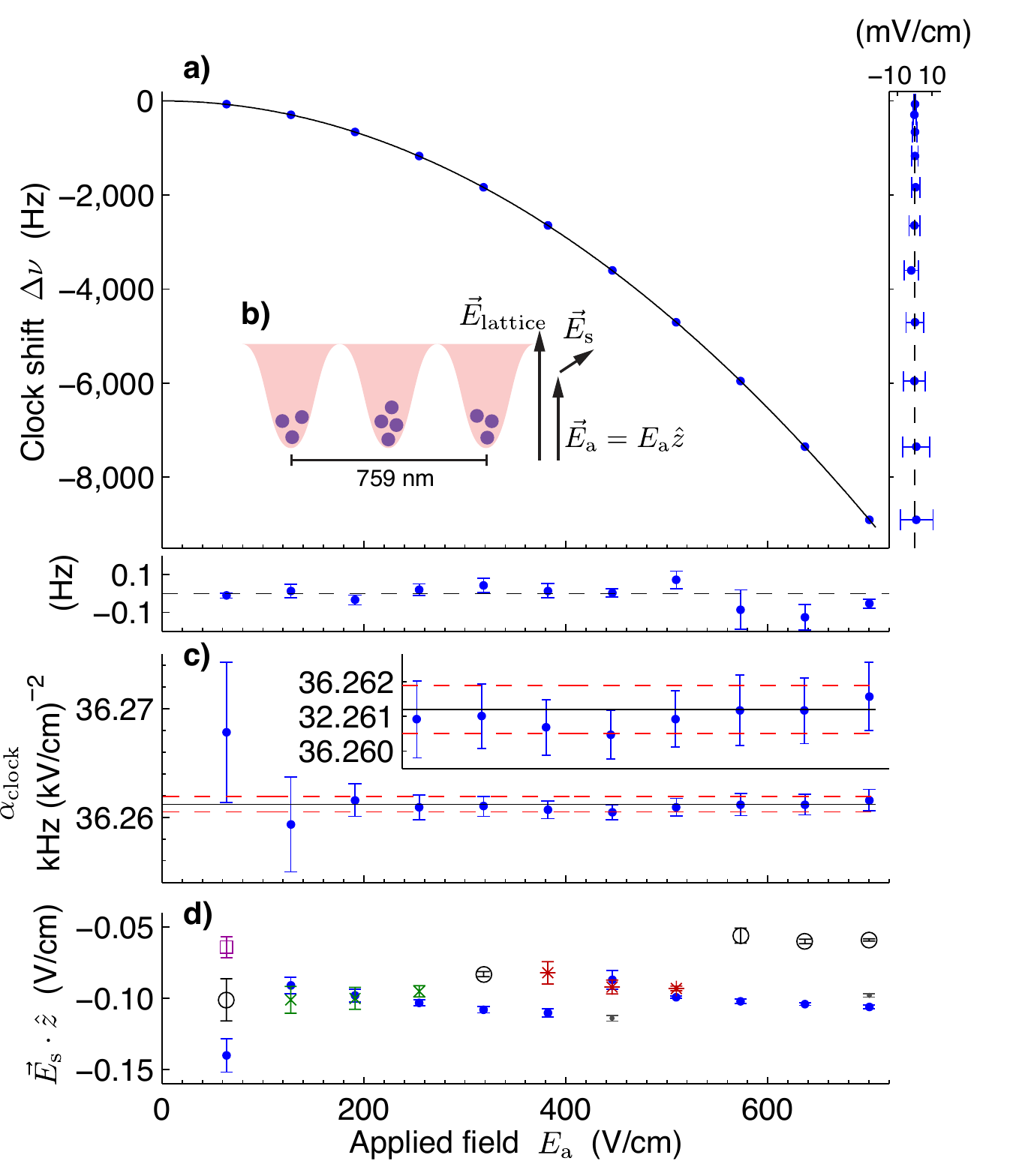}
\caption{The clock slows when stretched. a) Residuals of a quadratic fit $\Delta \nu = - \tfrac{1}{2} \alpha_\text{clock} E_\text{a}^2$ display measurement uncertainties in $\Delta \nu$ (below) and $E_\text{a}$ (right).  b) An inset depicts lattice trapped atoms and the relative orientations of $\vec{E}_\text{lattice}$, $\vec{E}_a$, and hypothetical static stray field $\vec{E}_s$.  c) At each $E_\text{a}$, we display $\alpha_\text{clock} = - 2\Delta \nu / E_\text{a}^2$ with combined statistic and systematic uncertainties (see Table~\ref{tab:systematics}). An inset shows data at higher resolution. Solid and dashed lines show the final result and standard error, respectively.  d) The component of a stray field $\vec{E}_\text{s} \parallel \vec{E}_\text{a}$ is precisely detected upon $\vec{E}_\text{a}$ reversal.  We observe more temporal variation in $E_\text{s}$ than correlation with $E_\text{a}$. Data with different marker styles were acquired on separate days.}
\label{fig:plots}
\end{figure}

Reversing $\vec{E}_\text{a}$~\cite{hunter1992precise} yields information about stray electric fields $\vec{E}_\text{s}$  parallel to $\vec{E}_\text{a}$ or differential contact potentials~\cite{fisher1976contact} (e.g.\ one electrode may develop a thin layer of ytterbium deposition).  The difference $\Delta \nu_B - \Delta \nu_C = 2 \alpha_\text{clock} \vec{E}_\text{a} \cdot \vec{E}_\text{s}$ reveals that $|\vec{E}_\text{s} \cdot \hat{z}|  \approx 0.1$~V/cm (see Fig.~\ref{fig:plots}); temporal drift and weak correlation with $\vec{E}_\text{a}$ are observed.  

Shifts due to a truly static field $\vec{E}_\text{s}$ subtract completely in Eq.~\ref{eq:starkShift}.  However, time dependent changes, notably those correlated with the polarity of $\vec{E}_\text{a}$, do not.  Because of careful experimental design, we are not aware of any appreciable stray field source with such a correlation.  Nevertheless, because an increase in dwell time $\tau_\text{v}$ potentially allows an accumulation of unknown stray charge (and thus a correlated stray field),  we varied $\tau_\text{v}$ over 0.8~s--12~s and resolved a small but negligible correlation in $\Delta \nu$ at applied fields twice the maximum used for reported data.  We observed no systematic variation in the measured polarizability $\alpha_\text{clock}$ with $E_\text{a}$.  We note that for data presented here, $E_\text{a}$ remained three orders of magnitude below the dielectric strength of fused silica, and five orders of magnitude below the characteristic level~\cite{lau1994electron} for ITO electron emission.  The time constant for electrode charging is $8~\mu$s.  Typically, 100 ms is allowed for settling.  Connecting high voltage to both electrodes creates an electric-gradient field;  by observing $\Delta \nu < 30$~mHz with 2~kV applied, we constrain the atoms' radial position to $\pm$10~mm, consistent with visual observations.



\begingroup
\squeezetable
\begin{table}
\caption{Uncertainty budget for a representative datum.  Errors in $E_\text{a}$ contribute twice the uncertainty as those in $\Delta \nu$ due to the dependence $\alpha_\text{clock} = -2 \Delta \nu / E_\text{a}^2$.  This factor of two is included in the tabulated quantities below. The total uncertainty in $\alpha_\text{clock}$, for the particular measurement shown, is found by summing all contributions in quadrature.}
\begin{ruledtabular}
\begin{tabular}{p{0.025 in}p{1.2in}p{0.2in}p{1.6in}}
\multicolumn{2}{c}{Uncertainty source}	  			& \multicolumn{1}{l}{$\times 10^{-6}$}	& \multicolumn{1}{c}{Notes/conditions}		\\ \hline \hline
\multicolumn{2}{l}{Shift statistical error}		& 8.3	    	&   $\Delta \nu = -3 603.77(3)$~Hz (1800~s averaging) 	\\
\multicolumn{2}{l}{Higher-order Stark shifts}	& 0.01													\\
\multicolumn{2}{l}{Electric field ($E_\text{a}$) errors:}	 	&		&  $E_\text{a} = 445.836(4)$~V/cm						\\
& Voltmeter systematic					& 16.4	&  Regulated $670.3966(55)$~V						\\
& $R_\text{leak}$ voltage division			& 0.1		&  $I_\text{leak} =  2.1$~nA; 20 k$\Omega$~leads  	         		\\
& Finite electrode size					& 1     	& Atoms centered $\pm$10~mm						\\
& Electrode parallelism					& 4 		& $\theta_\text{wedge} < 7~\mu$rad						\\
& Electrode deformation					& 0.8		& Warping of fused silica by gravity						\\
& Dielectric spacers						& 2		& Perturbation of ideal field due to three fused silica posts	\\
& Spacing $d$ (statistical)				& 1.6	     	& $N_\text{f} > 1700$ fringes spanned 							\\
& Spacing $d$ (systematic)				& 9 		& Fringe centering, wavemeter accuracy, stray etalons, stability		\\
& Etalon probe tilt, 	$\phi$				& 0.3		& $(1 - \cos \phi)$ error, $\phi \ll 0.5$~mrad; retro-coupling single-mode fiber  \\
& Yb thermal beam						& 0.06	& Dielectric $(\epsilon_r - 1) \sim 8 \times 10^{-9}$			\\ 
& Stray fields, static						& 0.04	& Uncertainty in $\vec{E}_\text{a}$ reversal 				\\
& Stray fields, varying					& 2		& $\Delta \nu$ correlation with $\tau_\text{v}$			\\
\hline
\multicolumn{2}{l}{Uncertainty in $\alpha_\text{clock}$}  & 21	&
\end{tabular}
\end{ruledtabular}
\label{tab:systematics}
\end{table}
\endgroup

Fig.~\ref{fig:plots}a shows the observed clock frequency shift quadratically as a function of $E_\text{a}$.  When fit to a polynomial, the data are consistent with no quartic, cubic, linear, or offset terms---an ideal demonstration of the Stark effect as non-degenerate perturbation theory.  No inhomogeneous line broadening is observed with increased shift, so the fractional statistical uncertainty in $\Delta \nu$ reduces as $E_\text{a}^{-2}$.  In contrast, the uncertainty of the applied voltage (the dominant systematic uncertainty) rises as $E_\text{a}^2$ according the specifications of our commercial voltmeters.  Fig.~\ref{fig:plots}c plots the polarizability inferred at each $E_\text{a}$.  Table~\ref{tab:systematics} lists the sources of measurement uncertainty at a particular applied field.  Taking the mean of all measurements, weighted by the total standard errors, we determine $\alpha_\text{clock} = 36.2612(7)$~kHz$(\text{kV/cm})^{-2}$. A least-squares functional fit (Fig.~\ref{fig:plots}a) yields a consistent value for $\alpha_\text{clock}$. Table~\ref{tab:results} demonstrates the agreement between this measurement and theoretical predictions.

\begin{table}[t!]
\caption{Comparison with theoretical predictions.  Results are also presented in SI and frequently used \emph{atomic units} (a.u.)~\cite{mitroy2010theory}.}
\begin{ruledtabular}
\begin{tabular}{lll|c}
\multicolumn{3}{c}{$\alpha_\text{clock} \equiv \alpha_e^{(0)} - \alpha_g^{(0)}$} &	\\
$[\,\text{kHz (kV/cm)}^{-2}]$ &	$[\text{ a.u.\ }$] & $10^{-39} [\text{C } \text{m}^2/\text{V}]$ 	& Reference \\ \hline \hline
40.1(3.7)		& 161(15)		& 2.65(25)							& \cite{dzuba2010dynamic}		\\
38.6(4.0)		& 155(16)		& 2.56(26)							& \cite{porsev2006multipolar}	\\
33(13)		& 134(51)		& 2.21(84)							& \cite{porsev1999electric}		\\ \hline
36.2612(7)	& 145.726(3)	& 2.40269(5)							& this work
\end{tabular}
\end{ruledtabular}
\label{tab:results}
\end{table}

Neither static nor BBR fields cause spin-magnitude-dependent ($\propto |m_F|^2$) \emph{tensor} Stark shifts because both clock states have insufficient total angular momentum ($F = F' = 1/2$)~\cite{angel1968hyperfine}.  Spin-sign-dependent ($\propto m_F$) \emph{vector} Stark shifts are absent as well:  BBR has no net polarization and static fields lack the time dependence to be circularly polarized~\cite{stalnaker2006dynamic}.  No opposite-parity states lie close to either clock state so no linear dependence of $\Delta \nu$ on $E_\text{a}$ is expected or observed.  A third-order effect~\cite{romalis1999zeeman} mixing the polarization due to the optical lattice ($E_\text{lattice} \approx 10$~kV/cm) and $E_\text{a}$ is expected to cause a $10^{-9}$ fractional error at the highest $E_\text{a}$.  A fourth-order term $\Delta \nu \propto E_\text{a}^4$ (the \emph{hyper-polarizability}) is responsible for a similarly sized effect.  We observed no systematic effect in $\Delta \nu$ upon varying the lattice intensity or polarization. We ensured that the atomic density did not systematically vary with application of $E_\text{a}$;  such a correlation could introduce contamination from the cold collision shift~\cite{lemke2011pwave}.  We resolved no shift by systematically varying the electrodes between grounded and floating configurations.

In an ideal blackbody environment at 300~K, the clock transition frequency is reduced from its value at $T=0$~K by $\Delta \nu_\text{BBR} = -1.273(1)$~Hz, where the uncertainty is now dominated by that of $\eta_\text{clock}$ (see Eq.~\ref{eq:pol}).  In practice, precise knowledge of $\alpha_\text{clock}$ and $\eta_\text{clock}$ is not sufficient to determine the effect of BBR, since knowledge of the environment is limited.  Non-uniformities arise due to temperature gradients, a hot effusive oven tip (850~K) and heated viewport (600~K), each about 30~cm away from the trapped atoms, and vacuum walls and viewports with less than unity emissivity and opacity.  An effective 1~K uncertainty in $T$ leads to a fractional uncertainty of $3 \times 10^{-17}$ in the room temperature clock.   A cryogenically shielded environment at $T=77(1)$~K with carefully controlled optical access can reduce the BBR shift uncertainty to the $1 \times 10^{-18}$ regime~\cite{middelmann2010tackling}.  A stray static field of 0.1~V/cm shifts the ytterbium clock transition -0.18 mHz, a fractional change of $4 \times 10^{-19}$.  A conductive enclosure at any temperature further ensures these, or smaller, stray static fields~\cite{lodewyck2011observation}.

We note that our present uncertainty in $\alpha_\text{clock}$ is competitive with the best known atomic or molecular polarizability, that of helium~\cite{schmidt2007polarizability}. Beyond timekeeping, possible metrological applications of the present work include high voltage measurement without the use of resistive dividers, and more directly, an atomic electric field meter sensitive at moderate field strengths.


\begin{acknowledgments}
This research was performed while J.A.S.\  held a National Research Council Research Associateship Award.  The authors acknowledge NIST and DARPA QuASaR for financial support, F.\ Calcagni and J.\ Perkins for contributing to electrode fabrication, T.\ Fortier and S.\ Diddams for femtosecond optical frequency comb measurements, and E.N.\ Fortson for a useful discussion on potential systematic effects.
\end{acknowledgments}

%

\end{document}